\documentclass[aps, prl, twocolumn, superscriptaddress, longbibliography]{revtex4-2}

\usepackage{graphicx}
\usepackage{hyperref}
\hypersetup{
 pdftitle = {Antithermalization in NV center},
 pdfauthor = {},
 breaklinks=true,
 pdfnewwindow=true,
 colorlinks=true,
 linktocpage=true,
 linkcolor={blue!45!green},
 citecolor={blue!45!red},
 filecolor=blue,
 urlcolor={blue!65!black}
}
\usepackage{xcolor}
\usepackage{float}
\usepackage{wrapfig}
\usepackage{tcolorbox}

\usepackage{mathtools, physics, amssymb, amsthm, bm, bbm}
\usepackage{dsfont, mathrsfs, yfonts, cancel, bbold, pifont}
\usepackage{tabularx}

\usepackage{soul, textcomp}
\usepackage[utf8]{inputenc}
\usepackage[english]{babel}
\usepackage[T1]{fontenc}

\usepackage{cleveref}

\let\tr\relax
\DeclareMathOperator{\tr}{Tr}

\let\ket\relax
\DeclarePairedDelimiter{\ket}{\lvert}{\rangle}
\let\bra\relax
\DeclarePairedDelimiter{\bra}{\langle}{\rvert}

\newcommand{\id}{\mathbbm{1}}

\newcommand{\erg}{\mathcal{E}}

\newcommand{\co}{\mathrm{c}}

\newcommand{\fr}{\mathrm{free}}
\newcommand{\eff}{\mathrm{eff}}

\newcommand{\iter}{\kappa}

\newfloat{extdata}{htbp}{loe}
\floatname{extdata}{Supplementary Fig.}
\crefname{extdata}{Extended Data Fig.}{Extended Data Figs.}

\newcommand{\cambridge}{Cavendish Laboratory, University of Cambridge, JJ Thomson Avenue, Cambridge CB3 0HE, United Kingdom}

\newcommand{\potsdam}{University of Potsdam, Institute of Physics and Astronomy, Karl-Liebknecht-Str. 24-25, 14476 Potsdam, Germany}

\begin{document}

\title{Ancilla mediated steady-state engineering in open quantum systems}

\author{Soham Pal\footnotemark[1]}
\thanks{These authors contributed equally}
\affiliation{\cambridge}

\author{Karen Hovhannisyan\footnotemark[1]}
\thanks{These authors contributed equally}
\affiliation{\potsdam}

\author{Omri Porat}
\affiliation{\cambridge}

\author{Shovan Dutta}
\affiliation{Raman Research Institute, Bengaluru 560080, India}

\author{Janet Anders}
\affiliation{\potsdam}
\affiliation{Department of Physics and Astronomy, University of Exeter, Exeter EX4 4QL, United Kingdom}

\author{Helena Knowles}
\email{hsk35@cam.ac.uk}
\affiliation{\cambridge}

\footnotetext[1]{These two authors contributed equally to this work.}

\begin{abstract}

Engineering the properties of a reservoir and its coupling to a quantum system is a powerful tool for simulating quantum thermodynamic processes and for generating otherwise inaccessible steady states. Yet tailoring both the reservoir and its coupling within a single platform remains challenging. Here we introduce a platform, in which an ancilla qubit mediates the coupling of a target system to a reservoir, providing independent control over the interaction form, coupling strength, and effective reservoir temperature. Our implementation uses the electron spin of a single nitrogen-vacancy center in diamond as the ancilla and a proximal $^{13}$C nuclear spin as the target. By alternating engineered unitary interactions with dissipative ancilla resets, we realize dynamics naturally described by a collision model, enabling straight-forward tracking of the work, heat, coherence, and entropy generated at every collision.  We experimentally demonstrate conventional thermalization and also realize anti-thermalization: the stabilization of the target system in a temperature opposite to that of its reservoir. Finally, harnessing this steady-state engineering, we utilize the nuclear spin as a quantum battery, achieving a steady-state ergotropy exceeding $70\%$ of the theoretical maximum.

\end{abstract}

\maketitle

\textit{Introduction}---
Coupling to an external environment is unavoidable for any real quantum system. However, this coupling need not be a nuisance: an appropriately engineered reservoir can itself be a resource \cite{Poyatos_1996, Koch_2016, Basilewitsch_2019}. Tuned system-bath interactions can be used to cool, to store and convert energy, and even to defy the natural direction of thermalization, stabilizing a target system in a population-inverted steady state instead of a Gibbs state at the bath temperature \cite{Uppalapati_2024}. In the language of quantum thermodynamics, such population inversion corresponds to an environment-assisted charging of a quantum battery \cite{Barra_2019, Hovhannisyan_2020, Quach_2020, Tacchino_2020, Morrone_2023b, Campaioli_2024}. Despite substantial theoretical interest in these control-intensive, non-equilibrium phenomena \cite{Vinjanampathy_2016, Cangemi_2024, Campaioli_2024}, an experimental platform combining coherent control, a genuinely tunable reservoir, and single-system readout has been missing, and \textit{anti-thermalization} has not yet been experimentally demonstrated.

In this work, we demonstrate how a single ancillary qubit and a tuneable temperature reservoir can be used to engineer novel open system thermodynamics and steady states in a target quantum system. Steady states are particularly important in areas like quantum information \cite{Harrington_2022, Schwarzhans_2026}, thermodynamic cooling \cite{Vinjanampathy_2016, Cangemi_2024} and energy storage \cite{Campaioli_2024}. We also develop a full theoretical description of our platform, allowing bookkeeping of thermodynamic quantities (entropy, work and heat), quantum correlations and mutual information between the ancillary qubit and the target quantum system at each stroke.
\begin{figure}[t!]
	\includegraphics[trim= 0cm 16.5cm 0cm 0cm, clip=true,width=0.9\columnwidth]{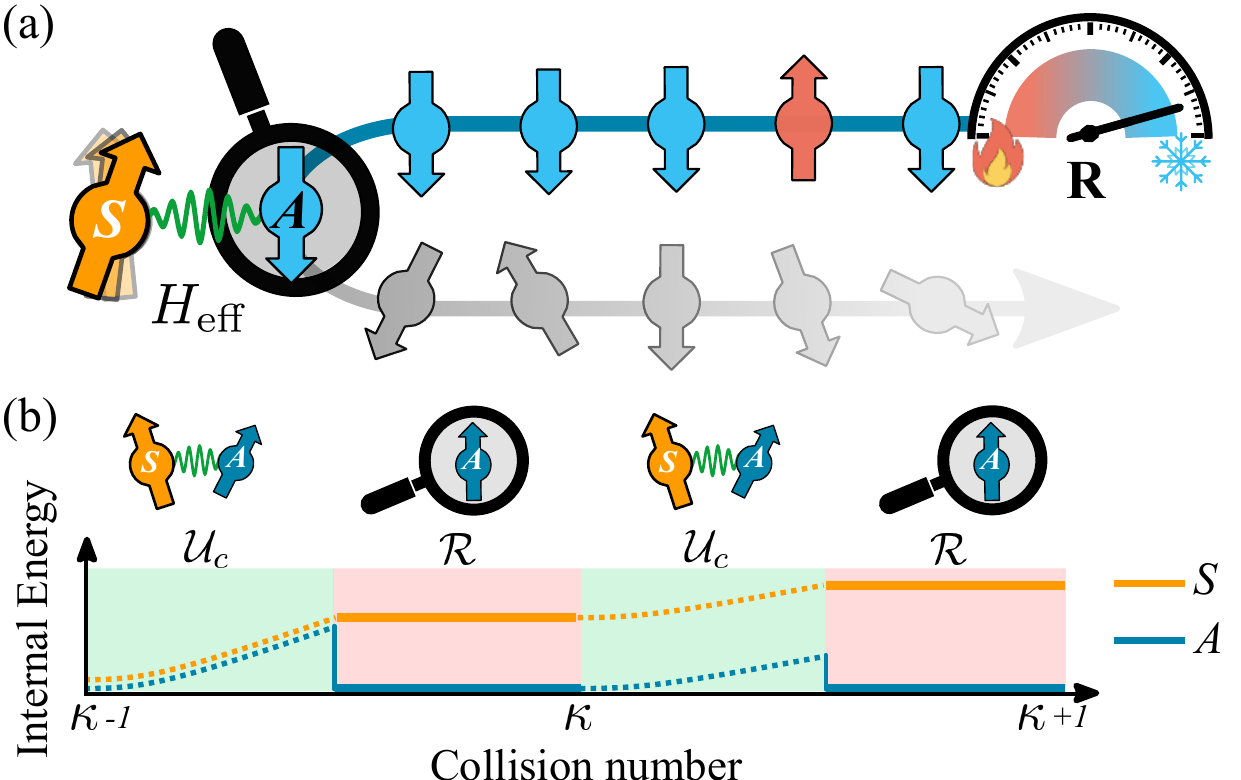}
	\caption{Schematic of the platform. (a) The system qubit $S$ interacts sequentially with a stream of ancilla spins, each initialized in either the $\ket{\uparrow}$ or $\ket{\downarrow}$ state, according to the reservoir spin temperature. (b) After each unitary interaction in the reversible $\mathcal{U}_\co$ stroke, the ancilla $A$ is reset to a chosen thermal state by the irreversible $\mathcal{R}$ stroke. The system–ancilla collision is subsequently repeated.
    }
    \label{fig:sketch}
\end{figure}
Our platform features (i) a coherently controllable quantum system, (ii) a reservoir of tunable effective temperature, and (iii) independent control over the form and strength of the system-reservoir coupling, including switching it on and off at will, Fig.~\ref{fig:sketch}. We implement this platform using solid-state spins \cite{Davies_1976, Loubser_1978, Jelezko_2004, Balasubramanian_2009, Mamin_2013, Awschalom_2018}, which offer long coherence times and coherent control \cite{Jelezko_2004, Maze_2008, Balasubramanian_2009, Weber_2010, Awschalom_2018, Schwartz_2018} making them a natural but so far under-exploited testbed for quantum thermodynamics \cite{Campbell_2026}, with only a handful of experiments to date \cite{Klatzow_2019, Hernandez-Gomez_2022, Hernandez-Gomez_2024, Niu_2024, Medina_2025, Kwon_2025}.

We demonstrate the versatility of our experimental platform by realizing two types of coupling between the ancilla $A$ (a single NV center in diamond) and the system qubit $S$ (a nearby ${}^{13}\mathrm{C}$ nuclear spin), which lead to different steady states. In particular, we report the experimental realization of \textit{anti-thermalization} of a quantum system, predicted in Ref.~\cite{Uppalapati_2024}, whereby sustained heat input from a positive-temperature reservoir drives the $^{13}$C nuclear spin to a negative-temperature steady state featuring population inversion. Once the steady state is reached, the heat flow stops automatically and the negative temperature is stable without requiring any energy pump (unlike a laser).
We then show that such a steady state acts as a quantum battery. We quantify its charge by measuring the incoherent ergotropy, reaching a steady-state value exceeding $70\%$ of the maximal extractable work after 30 collisions. Our work establishes ancilla-mediated collision dynamics as a powerful strategy for structured reservoir engineering in solid-state quantum systems.

\medskip

\textit{Tunable thermodynamics platform}---Our platform comprises of a quantum system $S$ coupled to an engineered reservoir $R$ through an intermediate ancilla qubit $A$. The ancilla mediates all system–reservoir interactions through repeated collision cycles. Each cycle proceeds in two strokes: reversible (coherent) and irreversible (dissipative). During the reversible stroke, the two-qubit system $S A$ undergoes a unitary evolution that can be engineered using coherent control to the qubit A \cite{Pal_2025}. The irreversible stroke re‑initializes the ancilla qubit to a prescribed reservoir spin temperature $T_R$. This mimics successive interaction with fresh reservoir qubits in a collision model, as schematically illustrated in Fig.~\ref{fig:sketch}.

At the start of collision $\iter$, the joint state is $\rho_S^{(\iter-1)} \otimes \tau_A$, where $\tau_A = p_\uparrow \dyad{\uparrow} + p_\downarrow \dyad{\downarrow}$ is a Gibbs state of $A$ with respect to its internal Hamiltonian $H_A = \omega_A \sigma^z_A/2$, at the reservoir temperature $T_R$. The reversible stroke realizes the unitary quantum channel \cite{Nielsen} $\mathcal{U}_\co[\bullet] \coloneq U_\co \bullet U_\co^\dagger$, where $U_\co$ is the engineered unitary. The resulting joint state,
\begin{align} \label{state_update}
    \rho_{SA}^{(\iter)} = \mathcal{U}_\co \big[\rho_S^{(\iter-1)} \otimes \tau_A \big],
\end{align}
determines the generated coherences and correlations between $S$ and $A$. Because the stroke is unitary, its entire energetic contribution is work,
\begin{align} \label{work}
    W^{(\iter)} = \tr\big[ \rho_{SA}^{(\iter)} \, H_{SA}^\fr \big] - \tr\big[ \rho_S^{(\iter-1)} \otimes \tau_A \, H_{SA}^\fr \big],
\end{align}
performed by the external MW drive. Here $H_{SA}^\fr = H_S \otimes \id_A + \id_S \otimes H_A + V_\mathrm{int}$ is the free Hamiltonian, with $H_S = \omega_S \sigma_S^z/2$ the internal Hamiltonian of $S$, and $V_\mathrm{int}$ the possible ``built in'' coupling between $S$ and $A$.

The irreversible stroke corresponds to a quantum channel that erases the $S$--$A$ correlations and resets $A$:
\begin{align} \label{reset_channel}
    \mathcal{R} \big[\rho_{SA}^{(\iter)}\big] \coloneq \tr_A \big[\rho_{SA}^{(\iter)} \big] \otimes \tau_A,  \quad \forall \rho_{SA}.
\end{align}
In practice, this operation is realized by a selective measurement of $\sigma^z_A$ (decorrelation) followed by a dissipative postprocessing step (thermalization); see Supplemental Material~\cite{SM}, Sec.~\ref{app:rand-to-av}. All energy change during the irreversible stroke is associated with heat
\begin{align} \label{heat}
    Q^{(\iter)} = \tr\big[ \rho_{SA}^{(\iter)} \, H_{SA}^\fr \big] - \tr\big[\rho_S^{(\iter)} \otimes \tau_A \, H_{SA}^\fr \big]
\end{align}
emitted into the reservoir.
We emphasize that, during this stroke, unlike $A$, the state of $S$ is not reset or forgotten. Therefore, as a result of the collision, its state is updated to $\rho_S^{(\iter)} = \tr_A \big[\rho_{SA}^{(\iter)}\big]$.

\begin{figure}[t!]
	\includegraphics[trim= 0cm 16cm 0cm 0cm, clip=true,width=0.8\columnwidth]{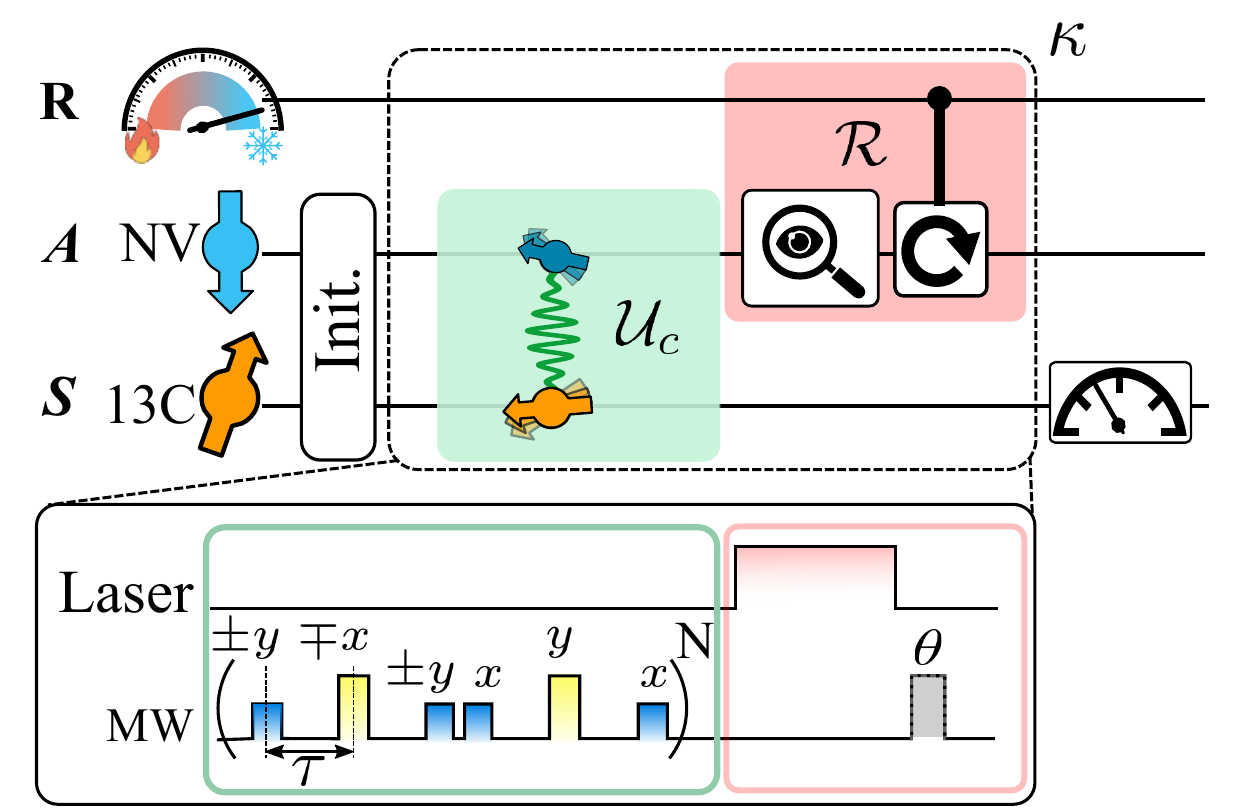}
	\caption{Circuit diagram for the solid-state implementation of the tunable thermodynamic platform with $\mathcal{U}_\co$ (green) and $\mathcal{R}$ (red) representing the unitary and dissipative strokes, respectively. The lower panel shows the pulse sequence used to generate $\mathcal{U}_\co$ and $\mathcal{R}$ strokes. The yellow (blue) bars are $\pi$ ($\pi/2$) pulses with phases labeled above and the gray pulse shows $\theta$ rotation, with $\tau$ as the inter-pulse delay.
	}
    \label{fig:platform_implementation}
\end{figure}

This platform thus naturally realizes a collision model of driven quantum open‑system dynamics~\cite{Rau_1969, Scarani_2002, Bruneau_2014, Cattaneo_2021, Prositto_2025, Ramon-Escandell_2025, Lacroix_2025}. It features energetically separated strokes while permitting controlled generation and destruction of quantum coherence and correlations across cycles. For different engineered interaction channels and reservoir temperatures $T_R$, we probe the resulting energy flows and the approach to the steady state for the system.

\medskip

\textit{Experimental realization of the platform}---Let us now map the theoretical ingredients of our tunable thermodynamic platform onto a concrete physical realization in a single NV center in diamond coupled to a nearby $^{13}$C nuclear spin. 
The NV electronic spin realizes the ancilla qubit $A$. The electronic ground state ($S=1$) possesses a zero-field splitting of $D=2.87\,\mathrm{GHz}$ between the $m_s=0$ and $m_s=\pm1$ sublevels. An external magnetic field $B_0=26\,\mathrm{mT}$ applied along the NV axis lifts the degeneracy of the $m_s=\pm1$ states~\cite{Gruber_1997, Jelezko_2002} and polarizes the host $^{14}$N nuclear spin via cross relaxation at the excited-state level anti-crossing (ESLAC)~\cite{gulka2021room}. By resonantly driving the $\ket{0} \!\rightarrow\! \ket{m_s=+1}$ transition, we isolate an effective two-level ancilla in the rotating frame, directly corresponding to the qubit $A$ in the theoretical model.
A $532\,\mathrm{nm}$ optical illumination for a few $\mathrm{\mu s}$ preferentially pumps population in the NV $m_s=0$ state through spin non-preserving decay, coupling the NV spin to a cold Markovian reservoir~\cite{Klatzow_2019}. This irreversible process prepares the ancilla in a thermal state $\tau_A$ with $p_{\downarrow}\sim 0.9$ at room temperature, simulating a low spin temperature reservoir. Crucially, this dissipative stroke not only resets the ancilla but also destroys any correlations involving the NV electronic spin, exactly mirroring the channel $\mathcal{R}$ [Eq.~\eqref{reset_channel}]. A hot (higher energy, negative spin temperature) reservoir is engineered by applying the same optical reset followed by a MW pulse of angle $\theta$, which transforms the ancilla population to any intermediate reservoir temperature. Thus, the simulated reservoir temperature $T_R$ is experimentally tunable, dictating the flow of \textit{heat} into or out of $A$ [Eq.~\eqref{heat}].

A strongly coupled $^{13}$C nuclear spin acts as the system qubit $S$. In the rotating frame of the MW drive, the Hamiltonian describing the nuclear spin qubit $S$ and the NV electron spin qubit $A$ takes the form
\begin{equation} \label{SysHam_exp}
    H(t) = \gamma_n B_0 \hat{I}^z \otimes \id_{\mathrm{NV}} + \mathbf{A}\!\cdot\!\hat{\mathbf{I}} \otimes \hat{S}^z + H_{MW}(t),
\end{equation}
where $\hat{I}^{x,y,z} \coloneq \tfrac{1}{2} \sigma_S^{x,y,z}$ and $\gamma_n$ are the nuclear spin operators and gyromagnetic ratio, respectively.  $\hat{S}_z$ is the corresponding NV spin operator. The middle term in Eq.~\eqref{SysHam_exp} is the hyperfine coupling which is precisely the built in coupling $V_\mathrm{int}$ appearing in $H_{SA}^{\rm free}$ [Eq.~\eqref{work}]. The hyperfine vector $\mathbf{A} = A_\perp \hat{\mathbf{e}}^x + A_\parallel \hat{\mathbf{e}}^z$ is experimentally identified using a dynamical decoupling sequence (XY8-$N$)~\cite{Gullion_1990, Bar-Gill_2013, Ryan_2010, Taminiau_2012}, yielding $A_{\parallel}=2.5\,\mathrm{MHz}$ and $A_{\perp}=0.4\,\mathrm{MHz}$ for the NV–$^{13}$C pair used here, see Supplemental Material~\cite{SM}, Sec.~\ref{app:experiment}.
Coherent MW control implements the reversible stroke $\mathcal{U}_c$ of Eq.~\eqref{state_update}. Arbitrary unitaries $U_\co$ are generated using shaped MW pulses $H_{MW}(t)=\omega_x(t)\hat{S}^x+\omega_y(t)\hat{S}^y$, with Gaussian envelopes of duration $16\,\mathrm{ns}$. These pulses generate the ancilla-system coherences and correlations that define the work contribution $W^{(\iter)}$ in Eq.~\eqref{work}. 
While the ancilla can be prepared at a tunable spin temperature using short laser pulses, the $^{13}$C nuclear spin remains unaffected and its spin populations can be read out using an ESR-like protocol~\cite{Pal_2025}. To reset (depolarize) the nuclear spin and reset (initialize) the electronic spin, we apply a long ($200\,\mu$s) laser pulse that simultaneously reads out the electronic spin. The nuclear spin depolarization occurs due to excited-state hyperfine interactions and cross relaxation~\cite{Gaebel_2006, Fuchs_2008, Jacques_2009, Gali_2009}. We exploit this mechanism to reproducibly prepare a depolarized thermal state $\rho_S^{(\iter-1)}$ of the system qubit $S$ at the start of each collision cycle.

\begin{figure}[t!]
	\includegraphics[trim= 0cm 11cm 0cm 0cm, clip=true,width=1\columnwidth]{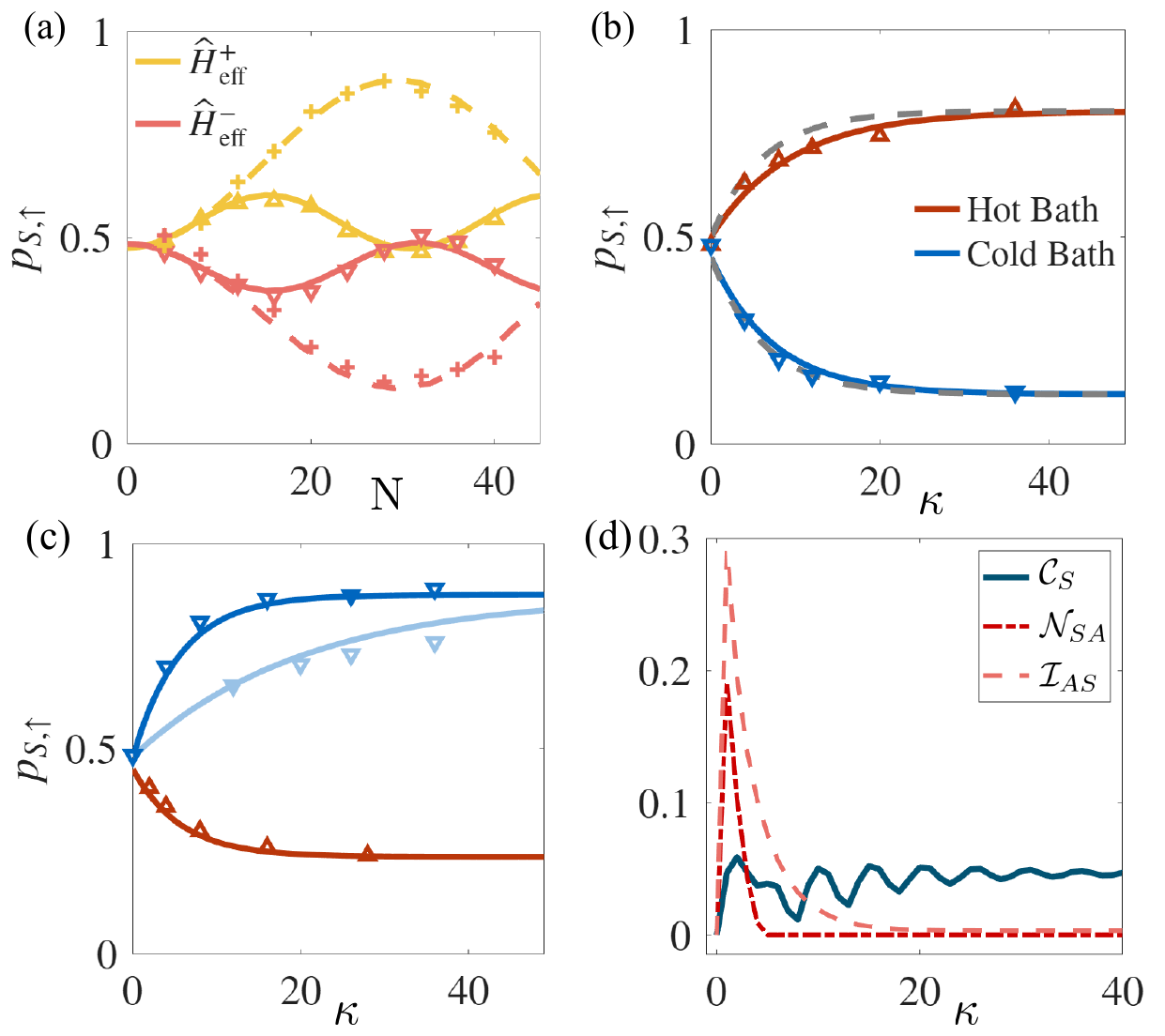}
	\caption{Simulated (solid/dashed) and experimental (markers) system energetics:  (a) Coherent evolution of the excited‑state population of $S$, $p_{S,\uparrow}$, under the reversible channels $H^+_{\mathrm{eff}}$ (yellow) and $H^-_{\mathrm{eff}}$ (red), shown for inter‑pulse delays of $376$ns (dashed) and $370$ns (solid) as a function of $N$ (see Fig.~\ref{fig:platform_implementation}).
    (b) Thermalisation of $S$ under $H^-_{\mathrm{eff}}$ for hot and cold baths at $370$ns (solid) and $376$ns (dashed) and N$=8$ as function of collision number, $\kappa$.  
    (c) Anti‑thermalisation under $H^+_{\mathrm{eff}}$, including a lighter trace for driving at $365$ns.  
    (d) Dynamics of system coherence $\mathcal{C}_{S}$ (solid) and quantum correlations (dashed), quantified via Negativity $\mathcal{N}_{AS}$ and Mutual Information $\mathcal{I}_{AS}$ for inter-pulse delay of $370$ns and N$=8$.
    }  
\label{fig:population_evolution}
\end{figure}

These elements provide a direct experimental implementation of the collision model described in the previous section, enabling cycle-by-cycle control of work, heat, coherence, and correlations in a solid-state quantum platform. Figure~\ref{fig:platform_implementation} shows the quantum circuit implementing  a single collision. The equilibration process is simulated by first preparing both the ancilla $A$ and system qubit $S$ at a known spin temperature, and then repeatedly applying the circuit for each collision $\kappa$. 

\medskip

\textit{Measurement driven thermalization and anti-thermalization}---To analyze how quantum correlations shape the steady state dynamics of the system qubit $S$ in the limit of many collisions, we focus on two distinct unitary channels: partial swap $U_\co^-$ (see Sec.~\ref{app:collision} of~\cite{SM} for explicit definition) and bit‑flipped partial swap
\begin{align}
U_\co^+ = (\id_S\!\otimes\!\sigma^x_A)\,U_\co^-\,(\id_S\!\otimes\!\sigma^x_A).
\end{align}
In the ideal collision model, these channels correspond to the effective Hamiltonians
\begin{align}\label{eff_Ham}
    H_\eff^\pm = \frac{i}{\tau_\co}\ln U_\co^\pm = \Omega\bigl(\sigma^+_S\!\otimes\!\sigma^\pm_A + \sigma^-_S\!\otimes\!\sigma^\mp_A\bigr),
\end{align}
where $\Omega$ is the ``coupling strength'' between $S$ and $A$. The two Hamiltonians act on different two‑qubit subspaces: $H_\eff^-$ couples $\ket{\downarrow\uparrow}$ and $\ket{\uparrow\downarrow}$, while $H_\eff^+$ couples $\ket{\downarrow\downarrow}$ and $\ket{\uparrow\uparrow}$, highlighting that the two channels differ only by a local flip on the ancilla.
We engineer these Hamiltonians using equally spaced $\pi$ and $\pi/2$ pulses with controlled phases~\cite{Schwartz_2018, Pal_2025}. Their effective strength $\Omega$ is set by the transverse hyperfine coupling and the filter function of the sequence~\cite{taminiau2014universal}. When the inter‑pulse delay, matches the nuclear‑spin resonance ($376\,\mathrm{ns}$ for the $^{13}$C--NV system we consider, see Sec.~\ref{app:experiment} of~\cite{SM}) we realize Eq.~\eqref{eff_Ham} and the coherent dynamics alone can drive the excited‑state population $p_{S,\uparrow}$ close to its extremal values for both $H_\eff^\pm$, as seen in Fig.~\ref{fig:population_evolution}(a) dashed lines and corresponding markers.

To isolate the role of the irreversible reset, we deliberately operate off resonance. Choosing an inter‑pulse delay of $370\,\mathrm{ns}$ suppresses coherent exchange such that the unitary stroke cannot, by itself, reach the extremal populations. The solid curves in Fig.~\ref{fig:population_evolution}(a) show that off‑resonant coherent dynamics remains bounded. Thus, in all equilibration experiments, any approach of $p_{S,\uparrow}$ toward its steady‑state value is necessarily driven by the reset stroke, which removes ancilla–system correlations and introduces dissipation. This makes the measurement driven nature of both thermalization and anti-thermalization explicit.

Under $H_\eff^-$, repeated collisions drive $S$ toward the ancilla state $\tau_A$~\cite{Scarani_2002}. Although the resulting temperature of $S$ need not equal $T_R$ when $\omega_S\neq\omega_A$, it always shares its sign, becoming literal thermalization when $\omega_S=\omega_A$. Figure~\ref{fig:population_evolution}(b) shows this behaviour: starting from a maximally mixed state, $S$ equilibrates to $p_{S,\uparrow}\approx0.1$ when A is coupled to a cold bath, and $p_{S,\uparrow}\approx0.8$ when A is coupled to the hot bath. The gray dashed line shows the same for an inter-pulse delay of $376$ ns.
For $H_\eff^+$, the collision map from Eq.~\eqref{state_update} and Eq.~\eqref{reset_channel} yields
\begin{align}\label{ideal_collisions}
    \rho_S^{(\iter)}=\sigma^x\tau_A \sigma^x+\cos(\Omega\tau_\co)^{2\iter}\bigl[\rho_S^{(0)}-\sigma^x\tau_A \sigma^x\bigr],
\end{align}
showing exponential convergence to the bit‑flipped thermal state $\sigma^x\tau_A \sigma^x$. This corresponds to \textit{anti‑thermalization}: the effective temperature of $S$ acquires the opposite sign of $T_R$. Figure~\ref{fig:population_evolution}(c) demonstrates this behaviour experimentally, with $p_{S,\uparrow}$ approaching $\approx0.9$ and $\approx0.2$ for $A$ coupled to a cold and hot reservoir, respectively. This population-inverted steady state demonstrates behavior opposite to conventional thermalization.
To present the tunability of the platform we also show the approach to the anti-thermal state for the cold bath at much slower rate by changing the inter-pulse delay further away from resonance, $365$ns. 

\begin{figure}[t!]
	\includegraphics[trim= 0cm 19cm 0cm 0cm, clip=true, width=1\columnwidth]{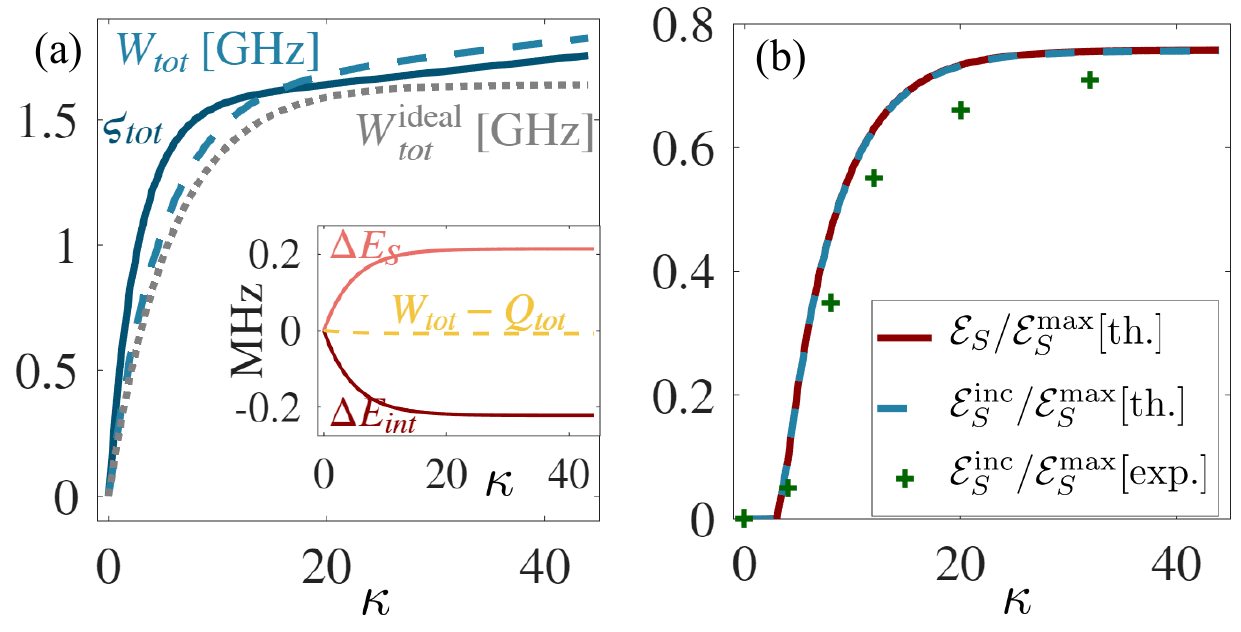}
	\caption{(a) Evolution of cumulative entropy ($\varsigma_\mathrm{tot}$) and work ($W_\mathrm{tot}$) injected as a function of the collision number $\kappa$. The continuous and dashed lines show the work and entropy production, respectively, for our experiment, whereas the dotted line shows the total work for the ideal $U_\co^+$, $W_\mathrm{tot}^{\mathrm{ideal}}$. Inset shows the energy balance of work and heat ($Q_\mathrm{tot}$) with the change in energy of the system.
    (b) Evolution of the full $\mathcal{E}_S$ and incoherent ergotropies $\mathcal{E}_S^{\mathrm{inc}}$ for a quantum battery realized by qubit $S$ and charged through qubit $A$ and reservoir $R$. The continuous curves are the simulated values, whereas the green crosses show the experimentally measured incoherent ergotropy.
	}
    \label{fig:collision}
\end{figure}

Figure~\ref{fig:population_evolution}(d) shows the coherence $\mathcal{C}_S^{(\iter)} = |\bra{\uparrow}\rho_S^{(\iter)}\ket{\downarrow}|$. The coherence remains small but finite asymptotically.
The nontrivial collisional dynamics shown in Fig.~\ref{fig:population_evolution}(c) requires correlations between $S$ and $A$ to be generated. Because $U_\co^\pm$ possess distinct nonzero entangling power \cite{Zanardi_2000, Kraus_2001, Scarani_2002}, $S$ and $A$ become entangled during the early collisions. As shown in Fig.~\ref{fig:population_evolution}(d), the negativity $\mathcal{N}_{SA}^{(\iter)}=\mathcal{N}(\rho_{SA}^{(\iter)})$ \cite{Nielsen} is appreciable for the first few iterations. We emphasize that vanishing entanglement does not imply absence of correlations. Towards this we also show the mutual information $\mathcal{I}_{AS}$ in Fig.~\ref{fig:population_evolution}(d).
The reset stroke, Eq.~\eqref{reset_channel}, erases these correlations and generates entropy, providing the ``\textit{arrow of time}'' that prevents oscillatory dynamics as seen in Fig.~\ref{fig:population_evolution}(a). The exact form of the entropy production per collision, $\varsigma^{(\iter)}$, can be found in Sec.~\ref{app:entropy_production} of~\cite{SM}. Figure~\ref{fig:collision}(a) shows the cumulative entropy production $\varsigma_\mathrm{tot}^{(\iter)}=\sum_{\nu=0}^\iter \varsigma^{(\nu)}$ along with the cumulative injected work and heat,
\begin{align*}
    W_\mathrm{tot}^{(\iter)} \coloneq \sum_{\nu=0}^\iter W^{(\nu)} \quad \mathrm{and} \quad Q_\mathrm{tot}^{(\iter)} \coloneq \sum_{\nu=0}^\iter Q^{(\nu)}.
\end{align*}
Note that $W_\mathrm{tot}^{(\iter)}$ and $Q_\mathrm{tot}^{(\iter)}$, being of the order of GHz, are much larger than the system energy scale, which is just a fraction of a MHz. However, the subtle energy balance is respected: Eqs.~\eqref{work} and~\eqref{heat} immediately yield
\begin{align} \label{subtle_balance}
    W_\mathrm{tot}^{(\iter)} - Q_\mathrm{tot}^{(\iter)} = \Delta E_S^{(\iter)} + \Delta E_\mathrm{int}^{(\iter)}.
\end{align}
Here $\Delta E_S^{(\iter)} = \tr(\rho_S^{(\iter)} H_S) - \tr(\rho_S^{(0)} H_S)$ and $\Delta E_\mathrm{int}^{(\iter)} = \tr(\rho_S^{(\iter)} \otimes \tau_A V_\mathrm{int}) - \tr(\rho_S^{(0)} \otimes \tau_A V_\mathrm{int})$ are the changes in, respectively, system and system--ancilla interaction energies by the end of collision $\iter$. These quantities are shown in the inset of Fig.~\ref{fig:collision}(a). In particular, we see that the internal energy of $S$ increases, following the population of its excited state [Fig.~\ref{fig:collision}(b)]. It is largely counterbalanced by the decrease in the energy stored in the hyperfine interaction, so that their sum matches the (much smaller) total energy input $W_\mathrm{tot} - Q_\mathrm{tot}$.

\medskip

\textit{A nuclear spin as a battery.}---The anti-thermalization of $S$ [Eq.~\eqref{ideal_collisions}] can be viewed as environment-assisted charging of a quantum battery \cite{Barra_2019, Hovhannisyan_2020, Quach_2020, Tacchino_2020, Morrone_2023b, Campaioli_2024}. The sustained population inversion places $S$ in an active, work-extraction state, so the nuclear spin acts as a charged quantum battery. When the pulse sequence is ``ideal'', namely, all spin rotations are exact and thus $U_\co^+$ is implemented with high precision, the asymptotic work cost of the collisions is essentially zero ($W^{(\iter\gg 1)}\approx 0$; see dotted line in Fig.~\ref{fig:collision}(a). Therefore, maintaining this charged state is, in principle, energetically free \cite{footnote_costs}, and moreover, the total work required to reach the steady charged state, $W_\mathrm{tot}^{\infty}$, is finite. Our antithermalizing ${}^{13}$C spin thus constitutes a proof-of-principle quantum battery, adding to the still modest set of experimental realizations \cite{Quach_2022, Hu_2022, Joshi_2022, Gemme_2022, MaillettedeBuyWenniger_2023, Niu_2024, Zhang_2024, Camposeo_2025, Song_2025, Li_2025, Hymas_2026, Li_2026}. Interestingly, as we establish numerically, finite work cost of fully charging the battery is the exception rather than the rule: $U_\co^\pm$ are the only collision unitaries for which total work and heat saturate in the large $\iter$ limit. For all other unitaries, $W_\mathrm{tot}^{(\iter)}$ and $Q_\mathrm{tot}^{(\iter)}$ diverge with $\iter$ even when $\rho_S^{(\iter)}$ relaxes to a fixed state. For example, the pulse sequence realized in practice has deviations from the ideal case, and therefore the actual total work and entropy production slowly, but eventually, diverge, as can be seen in Fig.~\ref{fig:collision}(a).

The relevant figure of merit of a battery is the ergotropy \cite{Allahverdyan_2004, Campaioli_2024}, which is the maximal work extractable by unitary operations: $\mathcal{E}_S = \tr(\rho_S H_S) - \min_U \tr(U\rho_S U^\dagger H_S)$.
For a qubit,
\begin{align}
    \erg_S = \hbar\omega_S \max\!\Big\{0,\; p_{S, \uparrow} - \tfrac{1}{2} + \sqrt{(p_{S, \uparrow} - \tfrac{1}{2})^2 + \mathcal{C}_S^2}\Big\},
\end{align}
with maximal value $\erg_S^{\max}=\hbar\omega_S$, attained for $\rho_S=\dyad{\uparrow}$. The quantity above is the ``full'' ergotropy, which depends nonlinearly on the coherence $\mathcal{C}_S$. When the coherence is unknown, designing the optimal extracting unitary becomes difficult, motivating interest in work extraction under limited state knowledge \cite{Safranek_2023, Chakraborty_2025, Hovhannisyan_2024b, Watanabe_2024}. In our setting, this challenge is avoided by considering the ``incoherent'' ergotropy, obtained from the dephased state $\mathrm{diag}(p_{S, \uparrow}, p_{S, \downarrow})$. For $p_{S, \uparrow} > 1/2$, the optimal unitary is simply $\sigma^x$ making the protocol experimentally straightforward and extracting
\begin{align}\label{erg_inc}
    \erg_S^{\mathrm{inc}} = \erg_S^{\max}\max\{0,\,2p_S-1\}.
\end{align}
Figure~\ref{fig:collision}(b) shows that $\erg_S^{(\iter)}$ and $\erg_S^{\mathrm{inc},(\iter)}$ remain close throughout the evolution, consistent with the small coherence generated in $S$ [Fig.~\ref{fig:population_evolution}(c)]. Since our measurements directly access the polarization of $S$, Eq.~\eqref{erg_inc} provides $\erg_S^{\mathrm{inc}}$ experimentally, and the obtained values closely follow the theoretical predictions [Fig.~\ref{fig:collision}(b)].

\medskip

\textit{Conclusions.}---We have introduced and experimentally realized an ancilla-mediated platform for engineering equilibrium steady states, implemented with a single NV-center electronic spin as a controllable ancilla for a strongly coupled $^{13}$C nuclear spin in diamond. Separating each collision cycle into a coherent work-injection stroke and a dissipative heat stroke lets us track work, heat, entanglement, and entropy production alongside the system's populations. By independently tuning the interaction Hamiltonian and the reservoir spin temperature, we realized both conventional thermalization and anti-thermalization: a phenomenon in which the system settles into a population inversion opposite in sign to that of its reservoir. We further showed that the antithermalized nuclear spin functions as a quantum battery, reaching a steady-state incoherent ergotropy exceeding 70\% of its maximal capacity after 30 collisions.
This platform naturally extends to ensembles of nuclear spins coupled to a single NV center, where some spins could serve as batteries powering other nuclear spins acting as thermal or information-processing elements---a route toward small autonomous quantum devices. More broadly, our results establish solid-state defect spins as a highly tunable testbed for quantum thermodynamics, particularly pointing to ancilla-mediated collision protocols as a general strategy for structured reservoir engineering.  Our results also open the door to multi-qubit batteries and tailored non-Markovian environments.

\bibliography{references}

\begin{thebibliography}{73}
\bibitem [{foo()}]{footnote_costs}%
  \BibitemOpen
  \href@noop {} {}\bibinfo {note} {Of course, this negligible work cost has
  only conceptual relevance here. The charge maintenance in this setup comes
  with a significant control cost, which in turn incurs macroscopic energetic
  input.}\BibitemShut {Stop}%
\end{thebibliography}


\clearpage
\onecolumngrid
\setcounter{section}{0}
\setcounter{equation}{0}
\setcounter{figure}{0}
\setcounter{table}{0}

\setcounter{secnumdepth}{3}

{\centering \Large\bfseries Supplemental Material\par}

\renewcommand{\thesection}{S\arabic{section}}
\renewcommand{\theequation}{S\arabic{equation}}
\renewcommand{\thefigure}{S\arabic{figure}}
\renewcommand{\thetable}{S\arabic{table}}

\section{Reset channel}
\label{app:rand-to-av}

The reset channel $\mathcal{R}$ [Eq.~\eqref{reset_channel} in the main text] is a two-step process. First, $A$ is measured in the eigenbasis $\{\ket{\downarrow}, \ket{\uparrow}\}$ of $\sigma^z_A$, and then a dissipative relaxation process induced by a laser drives its state to $\tau_A$. Namely, suppose we are at the end of the unitary stroke of collision $\iter$, so that the total system is in the state $\rho_{SA}^{(\iter)}$. Then, we selectively measure $\sigma^z_A$, collapsing the state of the system in the state
\begin{align}
    \rho_S^{({\iter})} &\otimes \ket{\downarrow}_{\! A \!} \bra{\downarrow} \quad \mathrm{with \;\; probability} \quad \bra{\downarrow} \rho_A^{(\iter)} \ket{\downarrow} ,
    \\
    \rho_S^{({\iter})} &\otimes \ket{\uparrow}_{\! A \!} \bra{\uparrow} \quad \mathrm{with \;\; probability} \quad \bra{\uparrow} \rho_A^{(\iter)} \ket{\uparrow} ,
\end{align}
where $\rho_A^{(\iter)} = \tr_S \big[ \rho_{SA}^{(\iter)} \big]$. The ensuing thermalization under the laser simply takes $A$, irrespective of whether it starts in $ \ket{\downarrow}_{\! A \!} \bra{\downarrow}$ or $ \ket{\uparrow}_{\! A \!} \bra{\uparrow}$ to $\tau_A$. As a result, by the end of this stroke, the total system ends up in the state $\rho_S^{(\iter)} \otimes \tau_A$ with probability $\bra{\downarrow} \rho_A^{(\iter)} \ket{\downarrow} + \bra{\uparrow} \rho_A^{(\iter)} \ket{\uparrow} = 1$, thus realizing the reset channel $\mathcal{R}$.

Note that, in reality, the hyperfine coupling between $S$ and $A$ will induce weak correlations between them during the thermalization process, so that the final state will not be an ideal tensor product. However, these correlations are weak, and therefore we ignore them throughout this paper.

\section{Ideal Unitary}
\label{app:collision}

The partial swap operation can be written as,
\begin{align}\label{SWAP}
\begin{split}
U_\co^- = &\, \tfrac{1+\cos(\Omega\tau_\co)}{2}\,\id_{SA}
          + \tfrac{1-\cos(\Omega\tau_\co)}{2}\,\sigma^z_S\!\otimes\!\sigma^z_A  \\
        & - i\sin(\Omega\tau_\co)\,(\sigma^+_S\!\otimes\!\sigma^-_A + \sigma^-_S\!\otimes\!\sigma^+_A),
\end{split}
\end{align}
where $\tau_\co$ is the duration of the coherent stroke and $\Omega$ sets the rotation angle. The second channel is a bit‑flipped partial swap,
\begin{align}
U_\co^+ = (\id_S\!\otimes\!\sigma^x_A)\,U_\co^-\,(\id_S\!\otimes\!\sigma^x_A).
\end{align}

\section{Details on the experiment}
\label{app:experiment}
\subsection{Sample}
The NV--\textsuperscript{13}C system investigated in this work resides in a Type~IIa single-crystal diamond with natural \textsuperscript{13}C abundance. The electronic-grade diamond, sourced from Element Six, was irradiated at the Leibniz Institute of Surface Engineering (IOM) using a 10~MeV electron accelerator (MB~10--30~MP, Mevex Corp., Stittsville, Canada). To reduce the incident electron energy to approximately 4--5~MeV, a 10~mm thick aluminum plate was inserted into the beam path.
Dosimetry was performed using a standard graphite calorimeter. The applied electron fluence was inferred from the measured dose using the known average stopping power for 5--10~MeV electrons, where a fluence of $10^{12}~\mathrm{cm^{-2}}$ corresponds to a dose of 360~Gy. In this work, the diamond was irradiated with a total dose of 2~kGy, corresponding to an electron fluence of $5.5\times10^{12}~\mathrm{cm^{-2}}$.

\subsection{Optical setup}
The experiments were performed using a home-built inverted confocal microscope operating at room temperature. A Nikon oil-immersion objective (NA = 1.4, 100$\times$) is mounted on a closed-loop $z$-piezo stage for precise axial positioning. Lateral scanning in the $xy$-plane is achieved using two orthogonally mounted galvanometric mirrors with an angular resolution of 15~$\mu$rad. To further improve the effective scan resolution to 3~$\mu$rad, we employ an asymmetric 4$f$ telescope (2$f_1$ + 2$f_2$) in the beam path. A static magnetic field is applied and aligned using two motorized rotation stages holding a neodymium permanent magnet. The alignment is optimized by monitoring the symmetry of the ODMR resonances around the zero-field splitting at 2.87~GHz.

\subsection{Microwave Control}
All pulsed measurements are performed using a Quantum Machines OPX+ integrated control and readout system, together with a 53~dB microwave amplifier from Mini-Circuits. The diamond is mounted on a No.~0 circular coverslip containing a coplanar waveguide with a 50~$\Omega$ termination for microwave delivery.
In all experiments, we use Gaussian-shaped microwave pulses with a fixed duration of $t_{\pi/2} = t_{\pi} = 16$~ns and a standard deviation of $t_{\pi}/5$. The $\pi$ pulses are applied at twice the amplitude of the $\pi/2$ pulses. Keeping the $\pi$ and $\pi/2$ pulses at equal duration ensures symmetric timing between pulses and minimizes timing errors. All pulses are resonant with the $\ket{0} \rightarrow \ket{+1}$ transition of the NV center.

\subsection{NV-$^{13}$C Hamiltonian calibration}
\begin{figure}[h!]
	\includegraphics[trim= 0cm 18cm 0cm 0cm, clip=true,width=0.8\columnwidth]{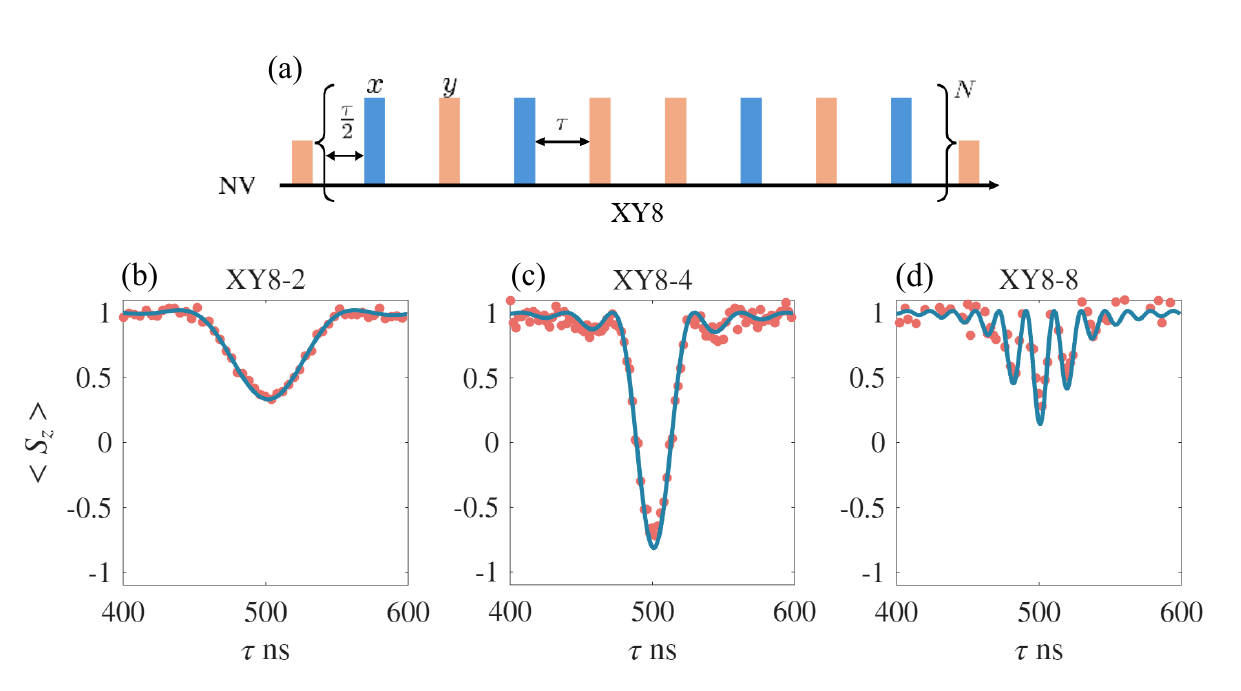}
	\caption{Hamiltonian calibration for NV-$^{13}$C spin register with XY8-N DD sequence. (a) Shows the XY8-N pulse sequence implemented on the NV spin. The short (tall) coloured rectangles represent $\frac{\pi}{2}$ ($\pi$) pulses and blue (orange) colour represent their phases (as shown), respectively. (b)-(d) show the simulated (solid blue line) and experimental (red markers) for the NV photo-luminescence (PL) as a function of inter-pulse delay, $\tau$ when implementing XY8-N with increasing order N. 
		\label{XYnv9}}
\end{figure}
We use the dynamical decoupling (DD) sequence called XY8 sequence \cite{gullion1990new} to characterise the coupling constants of the NV-$^{13}$C spin system, given in eq.1 of the main text. The conditional interaction between the NV-$^{13}$C can be probed by preparing the NV electronic spin in a superposition state with a $\frac{\pi}{2}$ pulse and subsequently applying a train of equally spaced, $\tau$, $\pi$ pulses in a specific order as shown in fig.~\ref{XYnv9}(a). When the inverse of the pulse spacing, $\tau$, is on resonance with the precession frequency of the nuclear spins coupled to electronic spin, their dynamics become correlated. As a result measurement on NV spin at these $\tau$ causes the NV PL to drop. This is shown in fig.~\ref{XYnv9}(b)-(d). increasing the number of $\pi$-pulse (order, N, of the XY8) results in higher correlations between NV-$^{13}$C, thus resulting in more drop in PL.  Here, $\tau = K\pi/\omega_I$ for odd integer $K$ ($K=3$ for fig.~\ref{XYnv9}). Here, the net nuclear precession frequency is given by $\omega_I = \sqrt{(\omega_L - A_z/2)^2 + (A_x/2)^2}$, with nuclear Larmor frequency $\omega_L = \gamma_n B_0$.
By fitting the experimental results with the simulations we find the parallel and the perpendicular component of the hyperfine coupling as shown in fig.~\ref{XYnv9}. All exp  were performed at 26mT and with 16ns $\pi$-pulse duration. 

\subsection{Generation of $H_{\mathrm{eff}}^+$}
\begin{figure}[h!]
	\includegraphics[trim= 0cm 22cm 0cm 0cm, clip=true,width=0.8\columnwidth]{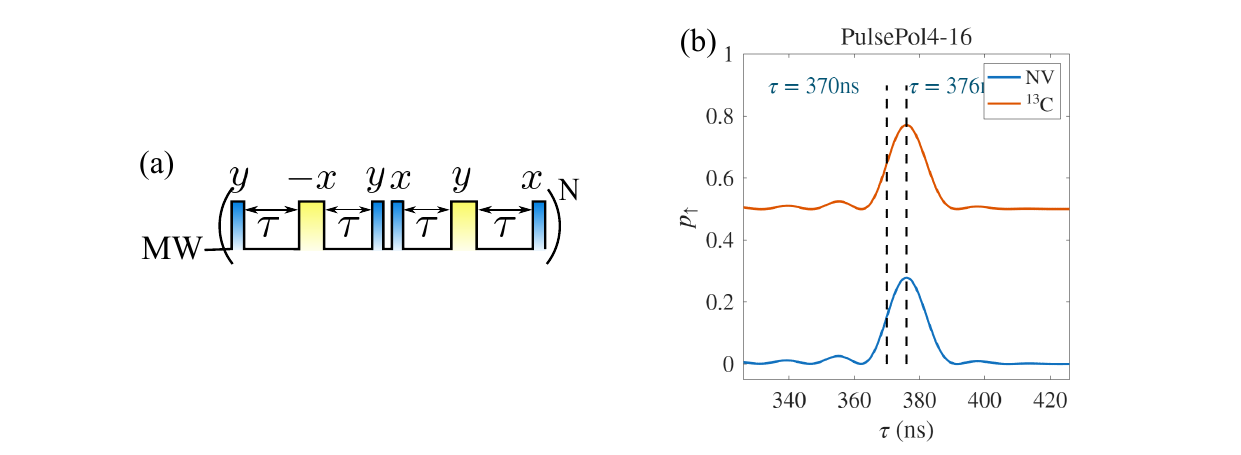}
	\caption{ (a) Pulse sequence used to generate $H_{\mathrm{eff}}^+$ in the main text. The blue (yellow) bars are the $\pi/2$ ($\pi$) pulses on the NV electronic spin. (b) Shows the simulated excited population dynamics for NV and $^{13}$C nuclear spin as function of inter-pulse delay.
		\label{PulsePolnv9}}
\end{figure}
Figure~\ref{PulsePolnv9}(b) shows what it means to be on resonant and off-resonant with the coupled nuclear spin while realising $H_{\mathrm{eff}}^+$ interaction in the main text. The delay between the pulses on NV, resonant between $\ket{0} \leftrightarrow \ket{+1}$ transition, $\tau$ for resonance is give by, 
\begin{align}
    \tau_k = \frac{k}{4\sqrt{(2\gamma_cB_0 - A_z)^2 + A_x^2}},
\end{align}
where $k = 1, 3, 5...$, and $\gamma_{C}$, $B_0$ and $A_{z(x)}$ are the carbon nuclear spin gyromagnetic ratio, externally applied magnetic field (26mT) and the z(x) component of the hyperfine coupling. All the simulation and experiments in the work have been done for $k=3$. The coupling values are mentioned in the main text.

\section{Entropy production during reset}
\label{app:entropy_production}

The reset stroke proceeds in two steps. First, one performs a projective measurement on $A$ in its energy eigenbasis $\{\ket{\downarrow}, \ket{\uparrow}\}$. Second, one lets $A$ thermalize with the (engineered) thermal reservoir at temperature $T_R$. Given the classical and macroscopic nature of the reservoir, we will assume it is an ideal bath.

Now, suppose we are at the end of the unitary stroke of the $\iter$'th collision, so that the system is in the state $\rho_{SA}^{(\iter)}$. Then, the first step of the reset stroke induces the state transformation
\begin{align}
    \rho_{SA}^{(\iter)} \; \longrightarrow \; \rho_S^{(\iter)} \otimes \mathcal{D}_{H_A}\big[\rho_A^{(\iter)}\big],
\end{align}
where the ``pinching map'' $\mathcal{D}_{H}[\rho]$ simply diagonalizes $\rho$ in the eigenbasis of $H$. The entropy produced in this step is merely the loss of mutual information between $S$ and $A$ \cite{Esposito_2010}, with the addition of entropy change of $A$ due to loss of coherence:
\begin{align}
    \varepsilon_1^{(\iter)} = S\big(\mathcal{D}_{H_A}\big[\rho_A^{(\iter)}\big]\big) + S\big(\rho_S^{(\iter)}\big) - S(\rho_{SA}^{(\iter)}).
\end{align}
During the second step---thermalization of $A$ with the reservoir $R$---entropy production is still given by the loss of mutual information, but this time between $A$ and $R$. The (very standard \cite{Esposito_2010}) argument is that $A$ and $R$ start in the product state
\begin{align} \label{ARinistate}
    \rho_{AR}^{(\iter)} = \mathcal{D}_{H_A}\big[ \rho_A^{(\iter)} \big] \otimes \tau_R,
\end{align}
where $\tau_R$ is a Gibbs state of $R$ at temperature $T_R$. They then undergo a unitary evolution $U_{AR}$ that thermlaizes $A$. Namely, the joint state evolves into
\begin{align} \label{ARevolution}
    \bar{\rho}_{AR}^{(\iter)} = U_{AR} \rho_{AR}^{(\iter)} U_{AR}^\dagger,
\end{align}
such that 
\begin{align} \label{zibil0}
\bar{\rho}_A^{(\iter)} = \tr_B[\bar{\rho}_{AR}^{(\iter)}] = \tau_A.
\end{align}
Thus, the entropy produced in this step is
\begin{align} \nonumber
    \varsigma_2^{(\iter)} &= S\big(\bar{\rho}_A^{(\iter)}\big) + S\big( \bar{\rho}_R^{(\iter)} \big) - S\big(\bar{\rho}_{AR}^{(\iter)}\big) 
    \\ \label{zibil1}
    &= S\big(\tau_A\big) + S\big( \bar{\rho}_R^{(\iter)} \big) - S\big(\rho_{AR}^{(\iter)}\big)
    \\ \label{zibil2}
    &= S\big(\tau_A\big) + S\big( \bar{\rho}_R^{(\iter)} \big) - S\big(\mathcal{D}_{H_A}\big[ \rho_A^{(\iter)} \big]\big) - S\big(\tau_R\big),
\end{align}
where, to obtain lines~\eqref{zibil1} and~\eqref{zibil2}, we used Eqs.~\eqref{zibil0}, ~\eqref{ARevolution}, and~\eqref{ARinistate}. Now, since we assume that $R$ is an ideal reservoir, and recalling that it receives $Q^{(\iter)}$ [Eq.~\eqref{heat}] amount of heat, we have that
\begin{align}
    S\big( \bar{\rho}_R^{(\iter)} \big) - S\big(\tau_R\big) = \frac{Q^{(\iter)}}{T_R}.
\end{align}
Plugging this into Eq.~\eqref{zibil2}, we thus get
\begin{align}
    \varsigma_2^{(\iter)} = \frac{Q^{(\iter)}}{T_R} + S(\tau_A) - S\big(\mathcal{D}_{H_A}\big[ \rho_A^{(\iter)} \big]\big).
\end{align}
Altogether,
\begin{align}
    \varsigma^{(\iter)} = \varsigma_1^{(\iter)} + \varsigma_2^{(\iter)} = \frac{Q^{(\iter)}}{T_R} + S(\tau_A) + S\big(\rho_S^{(\iter)}\big) - S(\rho_{SA}^{(\iter)}),
\end{align}
is the entropy produced during collision $\iter$.

\end{document}